\documentclass[aps,pra,twocolumn,superscriptaddress,showpacs,showkeys,amsmath,amssymb]{revtex4-1}
\usepackage{amsmath}
\usepackage{accents}
\usepackage{bm}
\usepackage[english]{babel}
\usepackage{graphicx}
\usepackage[usenames]{color}
\usepackage{colortbl}
\usepackage{mathtools}
\usepackage{commath}
\usepackage[font=footnotesize,figurewithin=none]{caption}
\usepackage{floatrow}
\usepackage{CJK}
\usepackage{multirow}
\usepackage{subfigure}
\usepackage[toc,page,title]{appendix}

\begin{document}

\title{Measuring distance between quantum states on a quantum computer}

\author{A. R. Kuzmak}
\email{andrijkuzmak@gmail.com}
\affiliation{Department for Theoretical Physics, Ivan Franko National University of Lviv,\\ 12 Drahomanov St., Lviv, UA-79005, Ukraine}

\date{\today}
\begin{abstract}
We propose protocols for determining the distances in Hilbert space between pure and mixed quantum states prepared on a quantum computer. In the case of pure quantum states, the protocol is based on measuring the
square of modulus of scalar product between certain states. Determination of the distance between mixed quantum states is reduced to measuring the squares of modules of scalar products between all pure states
included in the mixed states. In addition, we develop a protocol that allows one to determine the speed of evolution of the spin system simulated by a quantum computer. These protocols we apply to measure
distances and speeds of evolution of different quantum systems implemented on the ibmq-santiago quantum computer.
\end{abstract}

\maketitle

\section{Introduction}
\label{intro}

The concept of a distance between quantum states in Hilbert space \cite{Dodonov1999,Bengtsson2006,Tkachuk2011} has found its application in different fields of quantum mechanics related to the evolution
of quantum systems \cite{Anandan1990,Anandan1991,Abe1993,GNARQCS,brody2001,brach,brachass,brach1,OHfST,QBS1,ZNP1,QSGMBSD,QSLHUPOQC,Krynytskyi2019,Frydryszak2019,gqev3},
quantum entanglement \cite{brody2001,Shimony1995,Wei2003,Chen2014,Chen2017,GES,duan2001,zu2014,Entandgeom,torus,FMM,EQSGSSARII}, quantum computations \cite{OCGQC,GAQCLB,QCAG,QGDM,GQCQ}, etc.
It was shown that the distance, which the quantum system passes during the evolution in the Hilbert space, is related to the integral
of the uncertainty of energy that in turn defines the speed of evolution \cite{Anandan1990}. This distance is defined by the expression
\begin{eqnarray}
s=\int_0^{\tau}\sqrt{g_{tt}}dt,
\label{FSmetric}
\end{eqnarray}
where
\begin{eqnarray}
g_{tt}=\gamma^2\langle\psi(t)\vert\left(\Delta H\right)^2\vert\psi(t)\rangle,
\label{metrictensor}
\end{eqnarray}
and $\tau$ is a period of time. Here $\Delta H=H-\langle\psi(t)\vert H\vert\psi(t)\rangle$ is the energy unsertainty and $\vert\psi(t)\rangle=\exp{(-iHt)}\vert\psi_I\rangle$ is the state which the system described by the Hamiltonin $H$ achives during the time $t$
having started from the initial state $\vert\psi_I\rangle$, and $\gamma$ is a scale parameter.  We put $\hbar=1$, which means that the energy is measured in the frequency units. From equation (\ref{FSmetric}) follows that the speed of quantum evolution has the form
\begin{eqnarray}
v=\sqrt{g_{tt}}.
\label{speedevol}
\end{eqnarray}
This expression is called the Anandan-Aharonov relation \cite{Anandan1990}. The distance defined by expression (\ref{FSmetric}) is obtained from the Fubini-Study metric \cite{GNARQCS,Abe1992,Page1987,Kobayashi1969,Ozawa2018} for
two neighboring pure quantum states separated by an infinitesimal period of time. Indeed, the Fubini-Study distance \cite{Dodonov1999,Bengtsson2006,Tkachuk2011,Kobayashi1969,Bargmann1954} between two pure states
$\vert\psi_1\rangle$ and $\vert\psi_2\rangle$ is defined by the expression
\begin{eqnarray}
d^{FS}\left(\vert\psi_1\rangle,\vert\psi_2\rangle\right)=\gamma\sqrt{1-\vert\langle\psi_1\vert\psi_2\rangle\vert^2}.
\label{FSdistance}
\end{eqnarray}
Then, for two neighboring pure states $\vert\psi(t)\rangle$ and $\vert\psi(t+dt)\rangle$ separated by the period of time $dt$ the square of the Fubini-Study distance up to the second order in $dt$ takes form
\cite{Abe1992}
\begin{eqnarray}
ds^2=g_{tt}dt^2.
\label{FSnsdistance}
\end{eqnarray}
From this equation, it is easy to obtain an expression that allows one to define the distance (\ref{FSmetric}) that the system passes during the time $\tau$. Note that in some way equation (\ref{FSnsdistance})
can be derived from the Wootters distance \cite{Wootters1981}
\begin{eqnarray}
d^{W}\left(\vert\psi_1\rangle,\vert\psi_2\rangle\right)=\gamma\arccos\vert\langle\psi_1\vert\psi_2\rangle\vert,
\label{Wootersdistance}
\end{eqnarray}
minimal distance \cite{Pati1991}
\begin{eqnarray}
d^{min}\left(\vert\psi_1\rangle,\vert\psi_2\rangle\right)=\gamma\sqrt{2(1-\vert\langle\psi_1\vert\psi_2\rangle\vert)},
\label{mindistance}
\end{eqnarray}
or the definition of another distance between pure states (see, for instance, \cite{Dodonov1999,Ravicule1997}). In general, these distances are different, however, for
neighboring pure quantum states they coincide.

There are many definitions of the distance between mixed states in the physical literature: the Jauch-Misra-Gibson distance \cite{Jauch1968,Dieks1983}, the trace distance proposed by Hillery \cite{Hillery1987,Hillery1989},
the Bures-Uhlmann distance \cite{Bures1969,Uhlmann1976}, the Hilbert-Schmidt distance \cite{Dodonov1999,Bengtsson2006,Tkachuk2011,Anandan1991,Dieks1983,Baltz1990,Zyczkowski2001}. The most convenient for calculations
is the Hilbert-Schmidt distance. It is based on the Hilbert-Schmidt norm $\norm{A}_2\equiv\sqrt{{\rm Tr}\left(A^+A\right)}$. This distance between two mixed states $\rho_1$ and $\rho_2$ is defined as follows
\begin{eqnarray}
&&d^{HS}\left(\rho_1,\rho_2\right)=\gamma'\norm{\rho_1-\rho_2}_2=\gamma'\sqrt{{\rm Tr}\left(\rho_1-\rho_2\right)^2}\nonumber\\
&&=\gamma'\sqrt{{\rm Tr}\rho_1^2+{\rm Tr}\rho_2^2-2{\rm Tr}\rho_1\rho_2},
\label{Hilbert_Schmidtd}
\end{eqnarray}
where $\gamma'$ is a scale parameter. In the case of pure states the Hilbert-Schmidt distance turn into the Fubini-Study distance (\ref{FSdistance}) with $\gamma'=\gamma/\sqrt{2}$. It is important to note
that Hilbert-Schmidt distance is often used in quantum optics \cite{Dodonov1999,Knoll1995,Dodonov2003}.

In this paper, we propose protocols that allow one to determine the distance in Hilbert space between pure quantum states and define the speed of evolution of the quantum system prepared on a quantum computer
(Sec.~\ref{purestateprot}). Using this protocol, in Sec.~\ref{sec2} we obtain results for different quantum states and systems prepared on the ibmq-santiago quantum computer. Namely, we
measure the speed of evolution and distances between different states of spin-$1/2$ in the magnetic field (Subsec.~\ref{spinmagf}), the distance between the Schr\"odinger cat and factorized states (Subsec.~\ref{Schfact}),
and speed of evolution and distances between states achieved during the evolution of a spin-$1/2$ chain described by the Ising model (Subsec.~\ref{Isingmodel}). In addition, we develop and test a protocol which allows
measuring the distance between mixed quantum state prepared on a quantum computer (Sec.~\ref{mixedstate}). Conclusions are presented in Sec.~\ref{conc}.

\section{Protocol for determining the distance between pure quantum states \label{purestateprot}}

\begin{figure}[!!h]
\centerline{\includegraphics[scale=1.00, angle=0.0, clip]{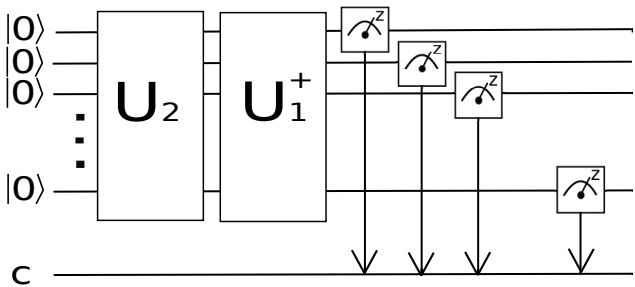}}
\caption{Quantum circuit for measuring the square of the modulus of scalar product between state $\vert\psi_1\rangle$ and $\vert\psi_2\rangle$ (\ref{sqmodsp}), which are performed by the $U_1$ and $U_2$ operators, respectively.}
\label{purestates}
\end{figure}

The Fubini-Study (\ref{FSdistance}), Wootters (\ref{Wootersdistance}) and minimal (\ref{mindistance}) distances contain the modulus of the scalar product of
states $\vert\psi_1\rangle$ and $\vert\psi_2\rangle$. The problem is to find a method that allows us to measure this modulus on a quantum computer. Let us represent these states as a transformation of the initial state
$\vert{\bf 0}\rangle=\vert 00\ldots 0\rangle$ under the action of the unitary operators $U_i$ as follows $\vert\psi_i\rangle=U_i\vert {\bf 0}\rangle$, where $\vert 0\rangle$ is the projection of the qubit on the positive
direction of the $z$-axis. We use such a representation because basically the initial state of quantum computers has the form $\vert {\bf 0}\rangle$. We obtain
\begin{eqnarray}
\vert\langle\psi_1\vert\psi_2\rangle\vert^2=\vert\langle{\bf 0}\vert U_1^+U_2\vert{\bf 0}\rangle\vert^2=\vert\langle{\bf 0}\vert\psi\rangle\vert^2,
\label{sqmodsp}
\end{eqnarray}
where $U_i^+$ is the conjugate transpose of $U_i$ and $\vert\psi\rangle=U_1^+U_2\vert{\bf 0}\rangle$. The problem of determination of the distance between pure quantum states $\vert\psi_1\rangle$ and $\vert\psi_2\rangle$
is reduced to measuring the probability corresponding to the reduction of $\vert\psi\rangle$ state into the $\vert{\bf 0}\rangle$ state. The protocol for measuring this probability is shown in Fig.~\ref{purestates}. Firstly we are
preparing the state $\vert\psi\rangle$ by applying the unitary operators $U_2$ and $U_1^+$ to the initial state $\vert{\bf 0}\rangle$, and then we make measurements of each qubit on $z$-axis.

\begin{figure}[!!h]
\centerline{\includegraphics[scale=1.00, angle=0.0, clip]{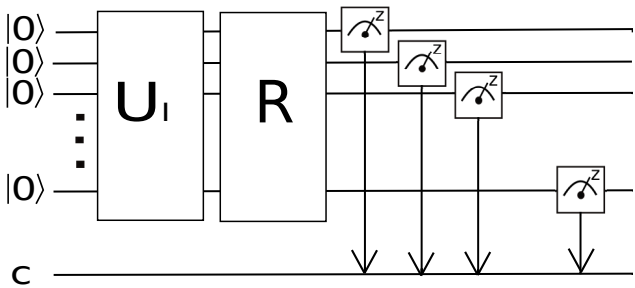}}
\caption{Quantum circuit for measuring the mean values (\ref{meanvalues2}). Operator $U_I$ provides preparation of the initial state $\vert\psi_I\rangle$. To provide the measurements
of each qubit on the basis $\vert 0\rangle$, $\vert 1\rangle$ of a quantum computer, the $U_{\alpha}$ operators should be expressed by the $\sigma^z$ operators. For this purpose
operator $R$ provides rotations of each qubit (\ref{transformationsofpauli}) so that the operator $U_{\alpha}$ transforms into the operator $\Sigma^z_{\alpha}$. Finally, the quantum computer
provides measurements of each qubit on the $z$-axis and the result is written into the classical register ${\rm c}$.}
\label{purestates2}
\end{figure}

We can also measure the speed of quantum evolution (\ref{speedevol}) on a quantum computer, which in turn allows us to determine the path that the system takes in Hilbert space during the period of time $\tau$ (\ref{FSmetric}).
The speed of evolution is defined by the component of metric tensor (\ref{metrictensor}). Using the fact that that the operator of evolution
$\exp{\left(-iHt\right)}$ and Hamiltonian $H$ mutually commute the $g_{tt}$ component can be rewritten in the following form
\begin{eqnarray}
g_{tt}=\gamma^2\left(\langle\psi_I\vert H^2\vert\psi_I\rangle-\langle\psi_I\vert H\vert\psi_I\rangle^2\right),
\label{metrictensorrew}
\end{eqnarray}
where $\vert\psi_I\rangle$ is the initial state prepared on a quantum computer by applying the certain unitary operator $U_I$ to the state $\vert{\bf 0}\rangle$ as follows $\vert\psi_I\rangle=U_I\vert{\bf 0}\rangle$.
As we can see, the mean values of $H$ and $H^2$ in the state $\vert\psi_I\rangle$ should be measured. For this purpose, we represent
the Hamiltonian in the form $H=\sum_{\alpha}h_{\alpha}U_{h_{\alpha}}$, where $h_{\alpha}$ are real parameters which determine the Hamiltonian and $U_{h_{\alpha}}$
are the Hermitian operators which satisfy the condition $U_{h_{\alpha}}^2=1$. It should be noted, that operators $U_{h_{\alpha}}$ determine the interactions in the system and they are represented by the basis gates
of a quantum computer. Then the mean values in equation (\ref{metrictensorrew}) take the form
\begin{eqnarray}
&&\langle\psi_I\vert H^2\vert\psi_I\rangle=\sum_{\alpha,\beta}h_{\alpha}h_{\beta}\langle\psi_I\vert U_{h_{\alpha}}U_{h_{\beta}}\vert\psi_I\rangle,\nonumber\\
&&\langle\psi_I\vert H\vert\psi_I\rangle=\sum_{\alpha}h_{\alpha}\langle\psi_I\vert U_{h_{\alpha}}\vert\psi_I\rangle,
\label{meanvalues}
\end{eqnarray}
The quantum computer provides the measurements of each qubit on the basis $\vert 0\rangle$, $\vert 1\rangle$
which consists of the eigenstates of $\sigma^z$ operators. This means that the $U_{h_{\alpha}}$ operators should be expressed by the $\sigma^z$ operators. For this purpose each of the qubits of the system
should be rotated as follows, if certain qubit $i$ in the term of Hamiltonian is defined by $\sigma^x_i$, $\sigma^y_i$ Pauli operator it should be rotated as follows
\begin{eqnarray}
\sigma^x=e^{-i\frac{\pi}{4}\sigma^y}\sigma^ze^{i\frac{\pi}{4}\sigma^y},\quad \sigma^y=e^{i\frac{\pi}{4}\sigma^x}\sigma^ze^{-i\frac{\pi}{4}\sigma^x}.
\label{transformationsofpauli}
\end{eqnarray}
As a result, mean values in equation (\ref{meanvalues}) take the form
\begin{eqnarray}
&&\langle\psi_I\vert U_{h_{\alpha}}U_{h_{\beta}}\vert\psi_I\rangle=\langle\tilde{\psi}_I^{R_{\alpha\beta}}\vert \Sigma^z_{\alpha}\Sigma^z_{\beta} \vert\tilde{\psi}_I^{R_{\alpha\beta}}\rangle,\nonumber\\
&&\langle\psi_I\vert U_{h_{\alpha}}\vert\psi_I\rangle=\langle\tilde{\psi}_I^{R_{\alpha}}\vert \Sigma^z_{\alpha}\vert\tilde{\psi}_I^{R_{\alpha}}\rangle,
\label{meanvalues2}
\end{eqnarray}
where $\vert\tilde{\psi}_I^{R_{\alpha\beta}}\rangle$, $\vert\tilde{\psi}_I^{R_{\alpha}}\rangle$ are the states reached from the state $\vert\psi_I\rangle$ by the rotation of certain qubits that the operators $U_{h_{\alpha}}$
transform into the operators $\Sigma^z_{\alpha}$ which consists of the compositions of the $\sigma^z_i$ Pauli operators. The method for the determination of mean values of the $\Sigma^z_{\alpha}$ operator is described in detail in papers
\cite{kuzmak2020,gnatenko2021,kuzmak2021}. This means the value in the state $\vert\psi\rangle$ is defined as follows
\begin{eqnarray}
\langle\psi\vert\Sigma^z_{\alpha}\vert\psi\rangle=p_+-p_-,
\label{meanvaluesSigmaz}
\end{eqnarray}
where $p_{\pm}$ are the probabilities that correspond to the mean values $\pm 1$. The protocol, which allows measuring the mean value, is shown in Fig.~\ref{purestates2}.

Let us apply these protocols to determine the distance between certain quantum states and values of the speed of evolution of some quantum systems.

\section{Determining the distance between pure quantum states on a quantum computer \label{sec2}}

\begin{figure}[!!h]
\centerline{\includegraphics[scale=0.50, angle=0.0, clip]{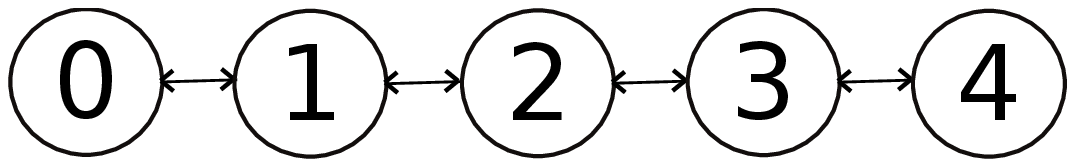}}
\caption{The ibmq-santiago quantum device consists of five superconducting qubits and has a linear structure. Each pair of qubits connected by the bidirectional arrow can be driven by the controlled-NOT operator
in a way that each of the qubits can be both a control and a target.}
\label{ibmq_santiago}
\end{figure}

In this section, we test our protocols on the ibmq-santiago quantum computer. We determine the distance between different pure quantum states prepared on this device. In addition, we simulate the time-evolution
and measure its speed in the case of a single spin in the magnetic field and spin system described by the Ising model. The ibmq-santiago is a 5-qubit quantum device designed by IBM (Fig.~\ref{ibmq_santiago}).
It can be used freely through the cloud service called the IBM Q Experience \cite{IBMQExp}. To perform the quantum circuits, this computer uses a controlled-NOT gate, the identity,
$R_z(\phi)$, $\sigma_x$ and $\sqrt{\sigma_x}$ single-qubit gates \cite{OpenQasm}. The $R_z(\phi)$ gate corresponds to rotating the qubit state around the $z$ axis by the angle $\phi$.
The ibmq-santiago allows us to prepare the $\sqrt{\sigma_x}$ quantum gate with fidelity $>99.6$\%, the controlled-NOT gate with fidelity $>93$\%. Almost all qubits are read with fidelity $>98$\%.

\subsection{Spin-1/2 in the magnetic field \label{spinmagf}}

The spin-$1/2$ in the magnetic field is defined by the Hamiltonian
\begin{eqnarray}
H=\frac{\omega}{2}{\bm \sigma}\cdot {\bf n},
\label{spin12ham}
\end{eqnarray}
where $\omega$ define the value of interaction between spin and magnetic field, ${\bm \sigma}=\sigma^x{\bf i}+\sigma^y{\bf j}+\sigma^z{\bf k}$ is the spin-$1/2$ operator, ${\bf n}=n_x{\bf i}+n_y{\bf j}+n_z{\bf k}$
is the unit vector which defines the direction of the magnetic field. Having started from the initial state $\vert\psi_I\rangle$ the evolution of such a system
can be expressed by the state vector \cite{Bengtsson2006,Tkachuk2011,TMTSPMF,TOSTSS}
\begin{eqnarray}
&&\vert\psi\rangle=e^{-i\frac{\omega t}{2}{\bm \sigma}\cdot {\bf n}}\vert\psi_I\rangle\nonumber\\
&&=\left[\cos\left(\frac{\omega t}{2}\right)I-i\sin\left(\frac{\omega t}{2}\right){\bm \sigma}\cdot {\bf n}\right]\vert\psi_I\rangle,
\label{spin12state}
\end{eqnarray}
where $I$ is an identity operator. Depending on the direction of the magnetic field and time of evolution the system can reach an arbitrary one-qubit state
\begin{eqnarray}
\vert\psi(\theta,\phi)\rangle=\cos\frac{\theta}{2}\vert 0\rangle+e^{i\phi}\sin\frac{\theta}{2}\vert 1\rangle,
\label{onequbit}
\end{eqnarray}
where $\theta\in[0,\pi]$, $\phi\in[0,2\pi]$ are some real parameters which determine the state. This state can be achieved from the initial state $\vert 0\rangle$ during the time $t$ in the case
of the perpendicular orientation of the magnetic field with respect to these states. Then, the parameters of state take the values $\theta=\omega t$ and $\phi=\arctan\left(\frac{n_y}{n_x}\right)-\frac{\pi}{2}$.
On the ibmq-santiago quantum computer, state (\ref{onequbit}) can be achieved by applying the $U(\theta,\phi,\lambda)$ gate to the state $\vert 0\rangle$, where $\lambda$ is a real parameter which can take the values from
the range $\lambda\in[0,2\pi]$. This gate is represented by the basis gates of the ibmq-santiago quantum computer as follows
\begin{eqnarray}
&&U(\theta,\phi,\lambda)\nonumber\\
&&=R_z(\phi+\pi)\sqrt{\sigma^x}R_z(\theta-\pi)\sqrt{\sigma^x}R_z(\lambda).
\label{U3gatebasis}
\end{eqnarray}
In the basis $\vert 0\rangle$, $\vert 1\rangle$ this gate reads
\begin{eqnarray}
U(\theta,\phi,\lambda)=\left( \begin{array}{ccccc}
\cos\frac{\theta}{2} & -e^{i\lambda}\sin{\frac{\theta} {2}} \\
e^{i\phi}\sin{\frac{\theta} {2}} & e^{i\left(\lambda+\phi\right)}\cos\frac{\theta}{2}
\end{array}\right).
\label{U3gate}
\end{eqnarray}

\begin{figure}[!!h]
\centerline{\includegraphics[scale=0.70, angle=0.0, clip]{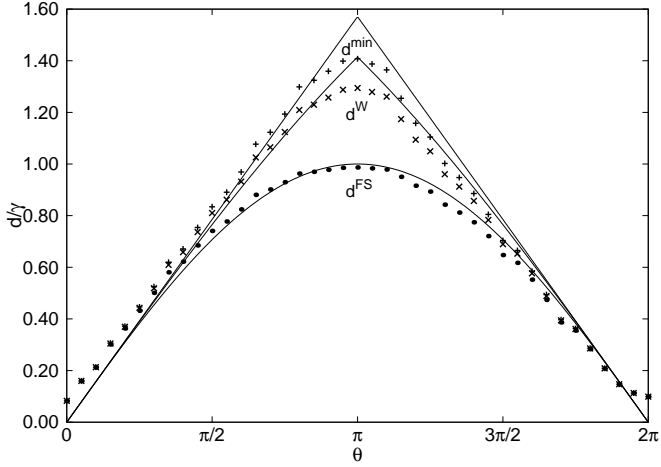}}
\caption{Dependencies of the Fubini-Study (\ref{FSdistance}), Wootters (\ref{Wootersdistance}) and minimal (\ref{mindistance}) distances between $\vert 0\rangle$ and (\ref{onequbit}) on parameter
$\theta$ obtained on the ibmq-santiago quantum computer. The solid lines show theoretical prediction.}
\label{dist_1qubit}
\end{figure}

Let us study the distance between two arbitrary quantum states of spin-$1/2$. Due to the symmetry properties of the state-space of spin-$1/2$, we can calculate the distance between $\vert 0\rangle$ and (\ref{onequbit}) states.
Recall those different definitions of distances (\ref{FSdistance}), (\ref{Wootersdistance}), (\ref{mindistance}) contain the modulus of scalar product between certain states. In our case of spin-$1/2$, the square of this value has the following form
\begin{eqnarray}
\vert\langle\psi_1\vert\psi_2\rangle\vert^2=\vert\langle 0\vert\psi(\theta,\phi)\rangle\vert^2=\cos^2\frac{\theta}{2}.
\label{scalarproduct}
\end{eqnarray}
This is the probability to measure state (\ref{onequbit}) on state $\vert 0\rangle$.
Using the fact that $\vert\psi(\theta,\phi)\rangle=U(\theta,\phi,\lambda)\vert 0\rangle$, the following value $\vert\langle 0\vert U(\theta,\phi,\lambda)\vert 0\rangle\vert^2$ should be measured.
Because probability (\ref{scalarproduct}) does not depend on parameters $\phi$ and $\lambda$, we set them to zero. Thus the protocol for determining probability (\ref{scalarproduct})
is shown in Fig.~\ref{purestates} with $U_1=I$ and $U_2=U(\theta,0,0)$. On the ibmq-santiago quantum computer, we measure this probability for different angles $\theta$, which changes in the range from 0 to $2\pi$ with
the step $\pi/20$. Here and further in the paper to obtain enough statistics, for each value we provide $1024$ measurements on the quantum computer. The results we substitute in expressions for distances (\ref{FSdistance}), (\ref{Wootersdistance}), (\ref{mindistance}).
The dependencies of distances on parameter $\theta$ are shown in Fig.~\ref{dist_1qubit}. Since we measure the distances in the case of one qubit, the results obtained on the quantum computer are in good agreement
with the theoretical prediction.

\begin{figure}[!!h]
\centerline{{\includegraphics[scale=0.70, angle=0.0, clip]{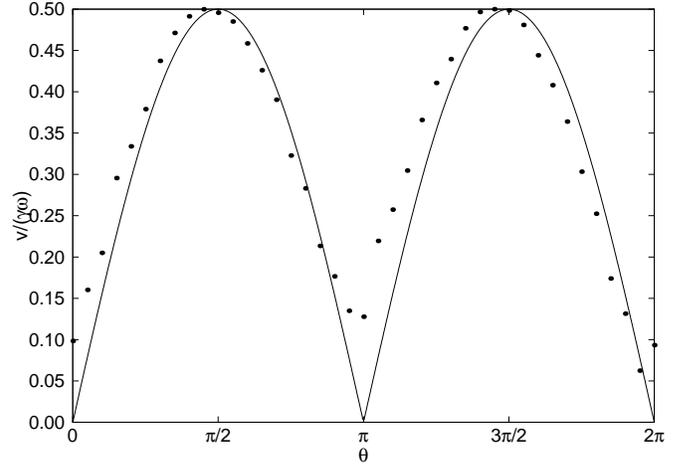}}}
\caption{Dependence of speed of evolution of spin-$1/2$ in the magnetic field on the angle between directions of magnetic field and vector of the initial state. The solid line shows the theoretical prediction
and the dots correspond to the value of speed obtained on the ibmq-santiago quantum computer.}
\label{vel_1qubit}
\end{figure}

Now let us study the speed of evolution of spin-$1/2$ in the magnetic field (\ref{spin12ham}). In this case $h_{\alpha}=\omega/2$, $U_{\alpha}={\bm \sigma}\cdot {\bf n}$ and we obtain
\begin{eqnarray}
&&\langle\psi_I\vert H^2\vert\psi_I\rangle=\frac{\omega^2}{4},\nonumber\\
&&\langle\psi_I\vert H\vert\psi_I\rangle=\frac{\omega}{2}\langle\psi_I\vert{\bm \sigma}\cdot {\bf n}\vert\psi_I\rangle,
\label{meanvaluesspin12}
\end{eqnarray}
As we can see, to determine the speed of evolution of spin-$1/2$ in the magnetic field only the mean value of the ${\bm \sigma}\cdot {\bf n}$ operator should be measured. For this purpose, we
rotate the qubit to direct the magnetic field along the $z$-axis which corresponds to the transformation of ${\bm \sigma}\cdot {\bf n}$ into $\sigma^z$. Then we can use expressions (\ref{meanvalues2}), (\ref{meanvaluesSigmaz})
with $\Sigma^z_{\alpha}=\sigma^z$ for determination of the mean value, where state $\vert\tilde{\psi}_I^{R_{\alpha}}\rangle$ has the form (\ref{onequbit}). Thus this mean value takes the form
\begin{eqnarray}
&&\langle\psi(\theta,\phi)\vert\sigma^z\vert\psi(\theta,\phi)\rangle\nonumber\\
&&= \vert\langle\psi(\theta,\phi)\vert 0\rangle\vert^2-\vert\langle\psi(\theta,\phi)\vert 1\rangle\vert^2.
\label{meanvaluesspin12}
\end{eqnarray}
Here we use the fact that $\sigma^z=\vert 0\rangle\langle 0\vert -\vert 1\rangle\langle 1\vert$. The problem reduces to the measurement of this mean value which should be substituted into (\ref{meanvaluesspin12}) and then into (\ref{metrictensorrew}) and (\ref{speedevol}). Making theoretical calculations
we obtain $\langle\psi(\theta,\phi)\vert\sigma^z\vert\psi(\theta,\phi)\rangle=\cos\theta$ and $g_{tt}=\gamma^2\omega^2\sin^2\theta/4$, $v=\gamma\omega\vert\sin\theta\vert/2$. On the ibmq-santiago we measure mean value
(\ref{meanvaluesspin12}) for different $\theta$. In Fig.~\ref{vel_1qubit} we show the results for the speed of evolution and compare them with theoretical predictions.

\subsection{Distance between Schr\"odinger cat and factorized states \label{Schfact}}

\begin{figure}[!!h]
\centerline{\includegraphics[scale=1.00, angle=0.0, clip]{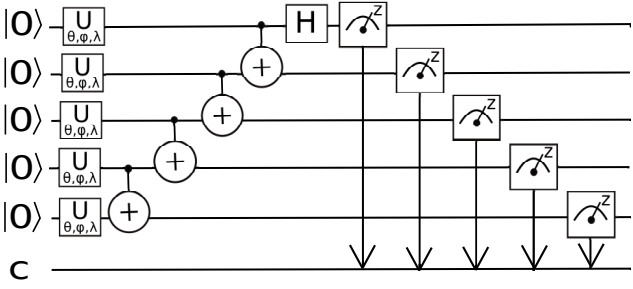}}
\caption{Quantum circuit for measuring the square of the modulus of the scalar product (\ref{sqmodsp}) between the five qubit Schr\"odinger cat state (\ref{schcatstate}) and the factorized state achieved by applying the $U(\theta,\phi,\lambda)$ gate to each qubit
in the $\vert 0\rangle$ state (\ref{factorizedstate}). To calculate this modulus, we conjure the Schr\"odinger cat state (\ref{schcatstate}). The Hadamar gate and a sequence of controlled-NOT operators define the conjugate transpose of the unitary operator which generates the Sch\"odinger cat state from the $\vert 00000\rangle$ state.
The quantum computer provides measurements of each qubit on the basis $\vert 0\rangle$, $\vert 1\rangle$ and the result is written into the classical register ${\rm c}$.}
\label{distSCHfact}
\end{figure}

In this subsection, we study the distance between 5-qubit states, namely, the Schr\"odinger cat state
\begin{eqnarray}
\vert\psi_{cat}\rangle=1/\sqrt{2}\left(\vert 00000\rangle+\vert 11111\rangle\right)
\label{schcatstate}
\end{eqnarray}
and factorized state
\begin{eqnarray}
\vert\psi_{fact}\rangle=\prod_{i=1}^5\vert\psi(\theta,\phi)\rangle_i,
\label{factorizedstate}
\end{eqnarray}
where $\vert\psi(\theta,\phi)\rangle_i$ is a single-qubit state defined by the expression (\ref{onequbit}).
Since the definition of distances between pure states (\ref{FSdistance}), (\ref{Wootersdistance}), (\ref{mindistance}) contains
the same modulus of the scalar products between certain states measured by a quantum computer, farther in the paper we study only the Fubini-Study distance (\ref{FSdistance}). This definition contains the square of the modulus
of scalar product between pure states which for the $\vert\psi_{cat}\rangle$ and $\vert\psi_{fact}\rangle$ takes the form
\begin{eqnarray}
&&\vert\langle\psi_{cat}\vert\psi_{fact}\rangle\vert^2\nonumber\\
&&=\frac{1}{2}\left(\cos^{10}\frac{\theta}{2}+\sin^{10}\frac{\theta}{2}+2\cos^5\frac{\theta}{2}\sin^5\frac{\theta}{2}\cos 5\phi\right).
\label{sqscalarproductSCHfact}
\end{eqnarray}
In Fig.~\ref{distSCHfact} we show the protocol for measuring this value on a quantum computer.

\begin{figure}[!!h]
\centerline{\subfigure[]{\label{dist_5qubit_theta}}{\includegraphics[scale=0.7, angle=0.0, clip]{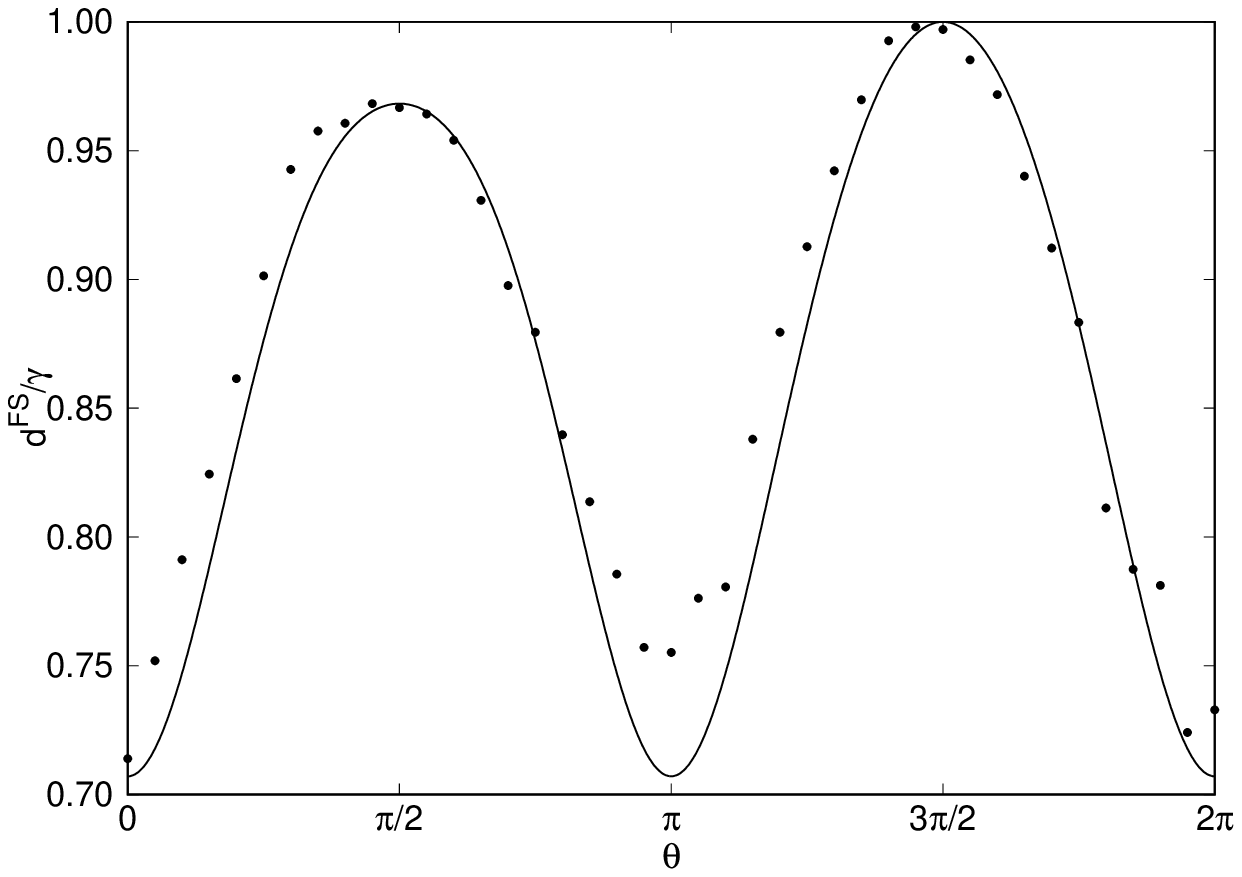}}}
\centerline{\subfigure[]{\label{dist_5qubit_phi}}{\includegraphics[scale=0.7, angle=0.0, clip]{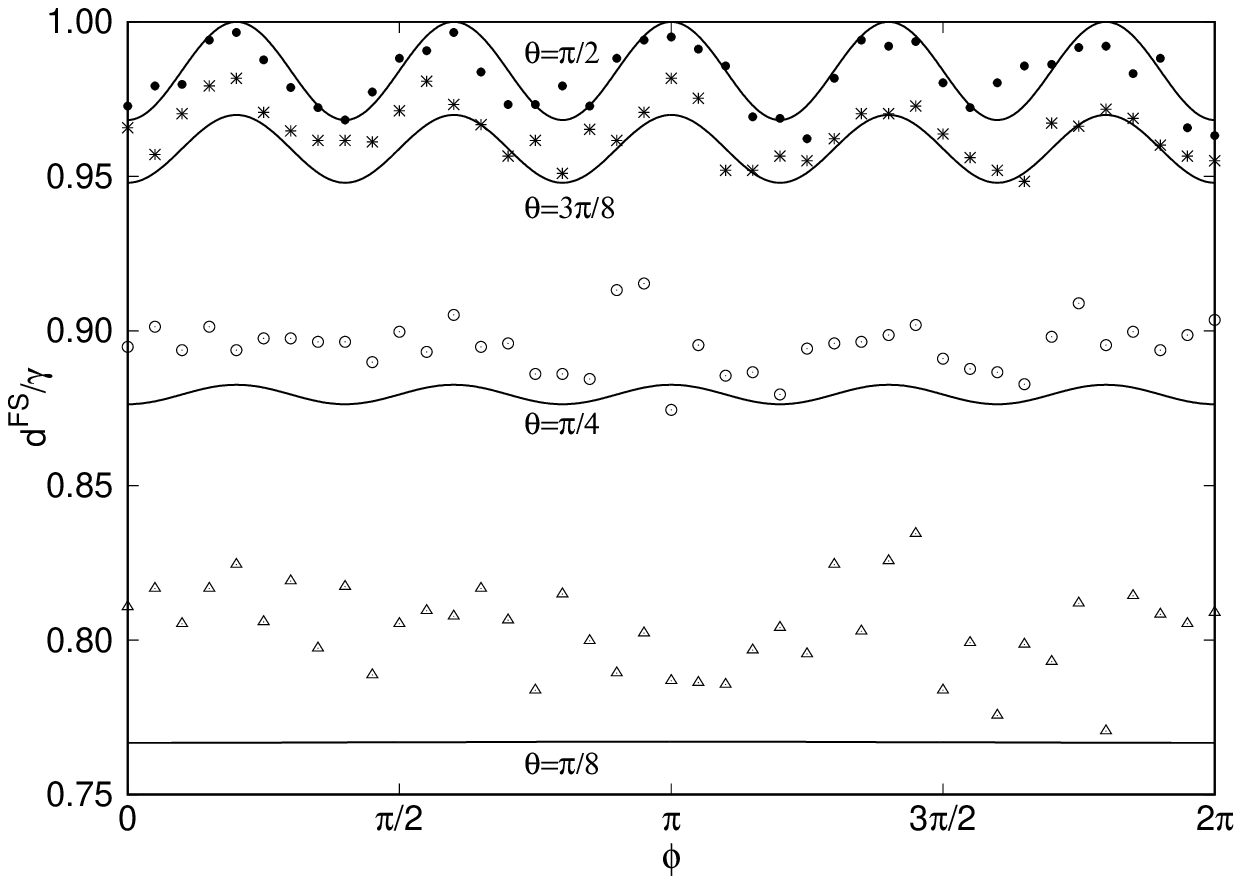}}}
\caption{Dependence of the Fubini-Study distance between Schr\"odinger cat (\ref{schcatstate}) and factorized (\ref{factorizedstate}) states on state parameters in the case of five qubits.
In figure (a) the dependence of the distance on angle $\theta$ in the case of $\phi=0$ is shown. In turn, in figure (b) the dependencies of the distances on angle $\phi$ for different
values of $\theta$ are represented. The solid lines show the theoretical prediction and the dots correspond to the results obtained on the ibmq-santiago quantum device.}
\label{dis5qubit}
\end{figure}

We take measurements for different value of $\theta$ in the case of $\phi=0$ (Fig.~\ref{dist_5qubit_theta}) and for different value of $\phi$ in the case of $\theta=\pi/2$, $3\pi/8$, $\pi/4$ and $\pi/8$ (Fig.~\ref{dist_5qubit_phi}).
We compare these dependencies with theoretical ones. As we can see, the closer the angle $\theta$ is to the value $\pi/2$ and $3\pi/2$, the more accurate the quantum computer performs the measurements. This fact follows the form
definition of the $U(\theta,\phi,\lambda)$ gate (\ref{U3gatebasis}). In the general case, this gate is performed by five basis operators. However, in the case of $\theta=\pm\pi/2$ the gate
is simplified as follows  $U(\pm\pi/2,\phi,\lambda)=R_z(\phi\pm\pi/2)\sqrt{\sigma^x}R_z(\lambda\mp\pi/2)$, which in turn reduced the error of this gate. Thus the preparation of the five-qubit factorized state $\vert\psi_{fact}\rangle$
with $\theta=\pi/2$, $3\pi/2$ requires ten basis operators less than in the general case.

\subsection{Ising model \label{Isingmodel}}

\begin{figure}[!!h]
\centerline{\includegraphics[scale=1.00, angle=0.0, clip]{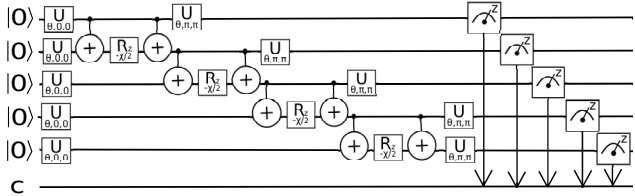}}
\caption{The quantum circuit allows one to measure the square of the modulus of the scalar product (\ref{sqmodsp}) between initial state (\ref{factorizedstate}) and state which is achieved during the evolution (\ref{IMevolution}).
To calculate this modulus, we conjure the state which is achieved during the evolution (\ref{IMevolution}). A set of operators $U (\theta,0,0)$ determine the initial state with $\phi=\lambda=0$,
the operator $U (\theta,\pi,\pi)$ is the conjugate transpose to the operator $U(\theta,0,0)$, and the unit consisting of two controlled-NOT operators
and $R_z(-\chi/2)$ generates the Ising interaction between certain spins. The quantum computer provides measurements of each qubit on the basis $\vert 0\rangle$, $\vert 1\rangle$ and the result is written into the classical register ${\rm c}$.}
\label{IM_dist_scheme}
\end{figure}

\begin{figure}[!!h]
\centerline{\includegraphics[scale=0.70, angle=0.0, clip]{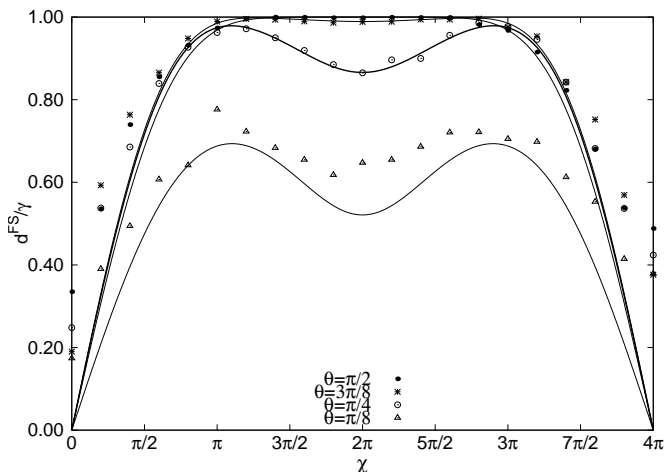}}
\caption{The Fubini-Study distance between initial state (\ref{factorizedstate}) with $\phi=0$ and different $\theta$ and state which is achieved during the evolution at the moment of time $t$ (\ref{IMevolution}).
The solid lines show the theoretical prediction and the dots correspond to the results obtained on the ibmq-santiago quantum device.}
\label{IM_dist_santiago}
\end{figure}

Based on the structure of the ibmq-santiago quantum device (see Fig.~\ref{ibmq_santiago}) the Ising model with the nearest neighbor interaction between spins can be simulated. We examine
the distance which separates the initial state and any other state achieved during the evolution of such a system. Hamiltonian of the Ising model with the nearest neighbor interaction has the form
\begin{eqnarray}
H=\frac{J}{4}\sum_{i=1}^{4}\sigma_i^z\sigma_{i+1}^z,
\label{IMham}
\end{eqnarray}
where $J$ defines the value of interaction between spins. Due to the fact that all terms in the Hamiltonian mutually commute, the evolution of this system can be expressed as follows
\begin{eqnarray}
&&\vert\psi(\chi,\theta,\phi)\rangle=e^{-iHt}\vert\psi_{fact}\rangle\nonumber\\
&&=\prod_{i=1}^4e^{-i\frac{\chi}{4}\sigma_i^z\sigma_{i+1}^z}\vert\psi_{fact}\rangle,
\label{IMevolution}
\end{eqnarray}
where $\chi=Jt$ is a parameter that depends on time and has a period of $4\pi$, $\vert\psi_{fact}\rangle$ is defined by expression (\ref{factorizedstate}). Each of the terms in the evolution operator
can be performed on a quantum computer using two controlled-NOT operators and one $R_z(-\chi/2)$ operator. Based on expression (\ref{sqmodsp}) in Fig.~\ref{IM_dist_scheme} we represent
a quantum circuit that allows measuring the square of the modulus of the scalar product between initial state (\ref{factorizedstate}) and state which is achieved during the evolution (\ref{IMevolution}). Because this value does not depend on $\phi$
\begin{eqnarray}
&&\vert\langle\psi_{fact}\vert\psi(\chi,\theta,\phi)\rangle\vert^2=\cos^8\frac{\chi}{4}+\sin^8\frac{\chi}{4}\cos^4\theta\nonumber\\
&&+\cos^4\frac{\chi}{4}\sin^4\frac{\chi}{4}\left(9\cos^8\theta+2\cos^6\theta-7\cos^4\theta+2\cos^2\theta\right)\nonumber\\
&&+\cos^6\frac{\chi}{4}\sin^2\frac{\chi}{4}\left(10\cos^4\theta-6\cos^2\theta\right)\nonumber\\
&&+\cos^2\frac{\chi}{4}\sin^6\frac{\chi}{4}\left(4\cos^8\theta+2\cos^6\theta-2\cos^4\theta\right)
\label{IMscalarprod}
\end{eqnarray}
we put $\phi=0$. On the ibmq-santiago quantum computer for different initial states, we measure the dependence of this value on parameter $\chi$. Using equation (\ref{FSdistance}) we calculate the Fubini-Study
distance between initial state (\ref{factorizedstate}) and state which is achieved during the evolution at the moment of time $t$ (\ref{IMevolution}). In Fig.~\ref{IM_dist_santiago} we demonstrate these results.
As in the previous case, the best coincidence of the results with the theoretical prediction we obtain for $\theta=\pi/2$.

\begin{figure}[!!h]
\centerline{\includegraphics[scale=0.70, angle=0.0, clip]{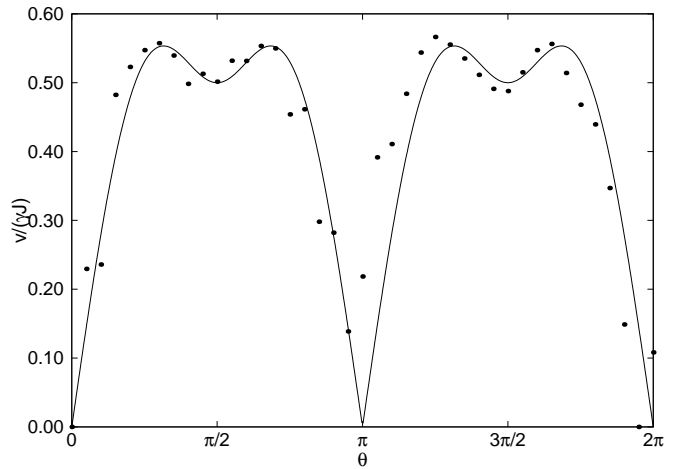}}
\caption{Dependence of the speed of evolution of the spin system described by the Ising model (\ref{IMham}) on the parameter $\theta$ which determines the initial state (\ref{factorizedstate}).
The solid line shows the theoretical prediction and the dots correspond to the results obtained on the ibmq-santiago quantum device.}
\label{IM_vel_santiago}
\end{figure}

Finally, substituting Hamiltonian (\ref{IMham}) with initial state (\ref{factorizedstate}) into expression (\ref{metrictensorrew}) we obtain
\begin{eqnarray}
&&g_{tt}=\frac{\gamma^2 J^2}{16}\left( 4+ \sum_{i\neq j=1}^4\langle\psi_{fact}\vert\sigma^z_i\sigma^z_{i+1}\sigma^z_j\sigma^z_{j+1}\vert\psi_{fact}\rangle \right.\nonumber\\
&&\left.-\left(\sum_{i=1}^4\langle\psi_{fact}\vert\sigma^z_i\sigma^z_{i+1}\vert\psi_{fact}\rangle\right)^2 \right)\nonumber\\
&&=\frac{\gamma^2 J^2}{16}\left(4+6\cos^2\theta-10\cos^4\theta\right).
\label{IMmetrictensorvel}
\end{eqnarray}
As we can see, to determine the speed of evolution the two- and four-spin correlation functions should be measured. Due to the structure of Hamiltonian (\ref{IMham}) the operators substituted into equations (\ref{meanvalues}) already
have the form $U_{h_{\alpha}}=\sigma_i^z\sigma_{i+1}^z$. To measure these correlations on a quantum computer we use expressions (\ref{meanvalues2}) and (\ref{meanvaluesSigmaz}) with $\vert\tilde{\psi}^{R_{\alpha}}_I\rangle=\vert\psi_{fact}\rangle$. Since the speed of evolution does not depend on parameter
$\phi$, on the ibmq-santiago device we obtain the dependence of the speed on parameter $\theta$ (Fig.~\ref{IM_vel_santiago}).

\section{Protocol for measuring the distance between mixed quantum states \label{mixedstate}}

In this section, we describe the protocol for determining the Hilbert-Schmidt distance between mixed quantum states (\ref{Hilbert_Schmidtd}) prepared on a quantum computer. An arbitrary mixed quantum state $\rho_i$ which
consists of a set of pure quantum states $\vert\psi_{\alpha}^{(i)}\rangle$ with appropriate weights $\omega_{\alpha}^{(i)}$ can be expressed as follows
\begin{eqnarray}
\rho_{i}=\sum_{\alpha}\omega_{\alpha}^{(i)}\vert\psi_{\alpha}^{(i)}\rangle\langle\psi_{\alpha}^{(i)}\vert,
\label{mixedstategen}
\end{eqnarray}
where $\sum_{\alpha}\omega_{\alpha}^{(i)}=1$. The traces included equation (\ref{Hilbert_Schmidtd}) can be represented as follows
\begin{eqnarray}
&&{\rm Tr}\rho_{i}^2=\sum_k\sum_{\alpha,\beta}\omega_{\alpha}^{(i)}\omega_{\beta}^{(i)}\langle k\vert\psi_{\alpha}^{(i)}\rangle\langle\psi_{\alpha}^{(i)}\vert\psi_{\beta}^{(i)}\rangle\langle\psi_{\beta}^{(i)}\vert k\rangle\nonumber\\
&&=\sum_{\alpha,\beta}\omega_{\alpha}^{(i)}\omega_{\beta}^{(i)}\langle\psi_{\alpha}^{(i)}\vert\psi_{\beta}^{(i)}\rangle\langle\psi_{\beta}^{(i)}\vert\left(\sum_k\vert k\rangle\langle k\vert\right)\vert\psi_{\alpha}^{(i)}\rangle\nonumber\\
&&=\sum_{\alpha,\beta}\omega_{\alpha}^{(i)}\omega_{\beta}^{(i)}\vert\langle\psi_{\alpha}^{(i)}\vert\psi_{\beta}^{(i)}\rangle\vert^2,\nonumber\\
&&{\rm Tr}\rho_{1}\rho_{2}=\sum_{\alpha,\beta}\omega_{\alpha}^{(1)}\omega_{\beta}^{(2)}\vert\langle\psi_{\alpha}^{(1)}\vert\psi_{\beta}^{(2)}\rangle\vert^2,
\label{traceinHSdist}
\end{eqnarray}
where $\vert k\rangle$ is a set of the basis vectors which defines the Hilbert space of the states $\rho_i$, and $\sum_{k}\vert k\rangle\langle k\vert$ is an identity operator defined in this space.
As we can see from (\ref{traceinHSdist}), the problem of determination of the distance between mixed quantum states is reduced to the problem of the determination of the squares of modules of scalar products between all pure quantum states
included by the mixed states. For this purpose the protocol described in section~\ref{purestateprot} is used. The results with appropriate products of weight factors are substituted into expressions (\ref{traceinHSdist})
and then into equation (\ref{Hilbert_Schmidtd}).

\begin{figure}[!!h]
\centerline{\includegraphics[scale=0.70, angle=0.0, clip]{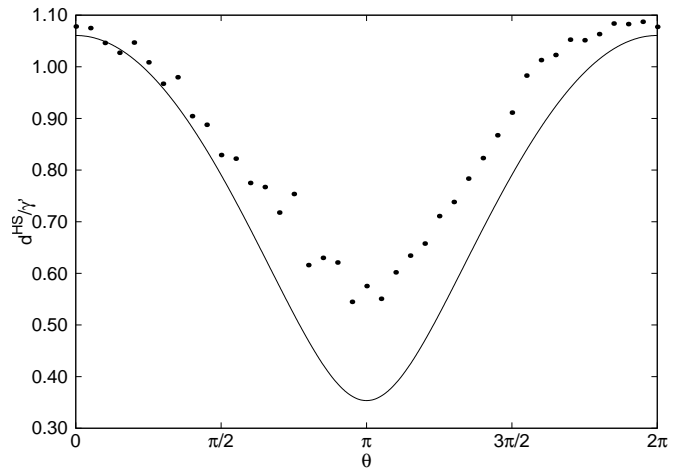}}
\caption{The dependence of the Hilbert-Schmidt distance between $\rho_1$ and $\rho_2$ states (\ref{mixedstateexample}) on state parameter $\theta$.
The solid line shows the theoretical prediction and the dots correspond to the results obtained on the ibmq-santiago quantum device.}
\label{dist_5qubit_mix}
\end{figure}

As an example, let us measure on the ibmq-santiago quantum computer the distance between the following quantum states
\begin{eqnarray}
&&\rho_1=\left(\cos\frac{\theta}{2}\vert 00000\rangle+\sin\frac{\theta}{2}\vert 11111\rangle\right)\nonumber\\
&&\left(\cos\frac{\theta}{2}\langle 00000\vert+\sin\frac{\theta}{2}\langle 11111\vert\right)\nonumber\\
&&\rho_2=\frac{1}{4}\vert 00000\rangle\langle 00000\vert+\frac{3}{4}\vert 11111\rangle\langle 11111\vert.
\label{mixedstateexample}
\end{eqnarray}
Thus, we want to define the distance between the pure state $\rho_1$ and mixed state $\rho_2$ consisting of
$\vert\psi_{1}^{(2)}\rangle=\vert 00000\rangle$, $\vert\psi_{2}^{(2)}\rangle=\vert 11111\rangle$ pure states with weight factors $\omega_1^{(2)}=1/4$, $\omega_2^{(2)}=3/4$, respectively.
Here the problem is reduced to the determination of all squares of modules between pure states $\cos\frac{\theta}{2}\vert 00000\rangle+\sin\frac{\theta}{2}\vert 11111\rangle$, $\vert 00000\rangle$, $\vert 11111\rangle$.
Using equation (\ref{Hilbert_Schmidtd}) for states (\ref{mixedstateexample}) we obtain
\begin{eqnarray}
d^{HS}\left(\rho_1,\rho_2\right)=\gamma'\sqrt{\frac{5}{8}+\frac{\cos\theta}{2}}.
\label{Hilbert_Schmidtd_example}
\end{eqnarray}
In Fig.~\ref{dist_5qubit_mix} we compare the results obtained on the ibmq-santiago quantum computer with theoretical ones. Since in the case of mixed quantum states we measure the squares of modules of scalar products
between all pure states included in these states, the errors accumulate from all measurements. In turn, this leads to the worse coincidence of the measurement results with the theory than in the case of pure states.

\section{Conclusions \label{conc}}

We have proposed the protocol that allows one to define the distance between pure states prepared on a quantum computer. To determine the distance between certain states the measurement results
on the initial state $\vert{\bf 0}\rangle$ of the quantum computer are enough to take. This fact makes our protocol easy to use and practical for calculations.
In addition, we have proposed the method for determining the speed of evolution of a quantum system simulated on a quantum computer. This method is based on measurement energy uncertainty which is included
in the well-known Anandan-Aharonov relation (\ref{speedevol}). The problem is reduced to the measurement of the mean values of spins and correlation functions of spins (\ref{meanvalues2}).
We have applied our methods to different pure quantum states and systems prepared on the ibmq-santiago quantum computer. As an example, we have determined the distances between pure states of spin-$1/2$
in the magnetic field. Depending on the direction of the magnetic field to the initial state we have measured the speed of evolution of such a system. We have also applied our
protocol to determine the distance between five-qubit pure states. Namely, we determine the distance between the Schr\"odinger cat state (\ref{schcatstate}) and factorized state (\ref{factorizedstate}).
We have also simulated the evolution of the system defined by the Ising Hamiltonian (\ref{IMham}). The distances between the initial state
and states achieved during the evolution have been measured. In addition, depending on the initial state the speed of evolution of such a system have been obtained.

Finally, we have developed the protocol to measure the distance between mixed quantum states prepared on a quantum computer. This protocol is based on the determination of the Hilbert-Schmidt norm (\ref{Hilbert_Schmidtd}).
We have shown that the distance between two mixed states is represented by the squares of the modules of scalar products between the pure quantum states included by the mixed states (\ref{mixedstateexample}).
We have applied this protocol to the determination of the distance between two states defined by density matrices (\ref{mixedstateexample}) prepared on the ibmq-santiago quantum computer.
Despite the fact that the measurements are performed for all possible scalar products between pure states, the experimental results are in good agreement with theoretical predictions.

\begin{acknowledgements}
This work was partly supported by Project 77/02.2020 (No.~0120U104801) from National Research Foundation of Ukraine.
We are grateful to Profs.  Volodymyr Tkachuk and Andrij Rovenchak for helpful advices.
\end{acknowledgements}

%



\end{document}